\input epsf.sty 
\input epsfig.sty 

\documentclass[12pt]{article}

\begin{document}

\thispagestyle{empty}
\renewcommand{\thefootnote}{\fnsymbol{footnote}}

\begin{flushright}
\end{flushright}

\vspace{.8cm}

\begin{center}
{\bf\Large   
Renormalization Group Reduction of the Henon Map and Application to the 
Transverse Betatron Motion in Cyclic Accelerators}\footnote{Supported 
by U.S. Department of Energy contract DE-AC02-76CH03073.} 

\vspace{1cm}

Stephan I. Tzenov and Ronald C. Davidson\\ 
{\it Plasma Physics Laboratory, Princeton University, Princeton, New Jersey,
08543}\\
\end{center}

\vfill

\begin{center}
{\bf\large   
Abstract }
\end{center}

\begin{quote}
The renormalization group method is applied to the study of discrete 
dynamical systems. As a particular example, the Henon map is considered 
as applied to describe the transverse betatron oscillations in a cyclic 
accelerator or storage ring possessing a FODO-cell structure with a 
single thin sextupole. A powerful renormalization group method is 
developed that is valid correct to fourth order in the perturbation 
amplitude, and a technique for resolving the resonance structure of the 
Henon map is also presented. This calculation represents the first 
successful application of a renormalization group method to the study 
of discrete dynamical system in a unified manner capable of reducing 
the dynamics of the system both far from and close to resonances, thus 
preserving the symplectic symmetry of the original map. 
\end{quote}

\vfill

\newpage


%
\pagestyle{plain}

\renewcommand{\theequation}{\thesection.\arabic{equation}}

\setcounter{equation}{0}

\section{Introduction and Basic Equations} 

It is increasingly important to develop an improved theoretical 
understanding of the nonlinear dynamics of charged particle beams in 
high energy accelerators and storage rings 
\cite{Chao,Syphers,Davidson,Bryant,Chatto1,Chatto2}. An individual 
particle propagating in an accelerator experiences a growth of amplitude 
of betatron oscillations in a plane transverse to the particle orbit 
whenever a perturbing force acts on it. This force may be of various 
origins, for example, high-order multipole magnetic field errors, 
space-charge forces, beam-beam interaction force, power supply ripples, 
or other external and collective forces. 

There is a growing number of robust analytical methods used to study 
the effects of the nonlinear behavior of beams in accelerators and 
storage rings, ranging from classical perturbation theory 
\cite{Lichtenberg,Nayfeh} to the Lie algebraic approach 
\cite{Dragt,Giacaglia,Cary}. The recently developed renormalization 
group (RG) method has been successfully applied to both continuous 
dynamical systems \cite{Chen,Oono,Kunihiro} and maps \cite{Goto,Kunihir} 
that are of general interest in the physics of accelerators and beams. 
The advantage of the RG method is associated with the fact that it is 
equally powerful to study finite-dimensional, as well as continuous 
systems. Therefore, it is also useful when applied to analyze the 
properties of chaotic dynamical systems in both the stability region and 
the globally stochastic region in phase space \cite{Arnold,Zwanzig,Tzenov}. 

While the RG method is well established in applications to continuous 
dynamical systems, the present paper demonstrates that the RG method 
can also be applied successfully to study discrete dynamical systems. 
As a particular example, we consider the Henon map \cite{Henon1,Henon2} 
as applied to describe the transverse betatron oscillations in a cyclic 
accelerator or storage ring possessing a FODO-cell structure with a 
single thin sextupole. The basic equations and Henon transfer map 
used in the present analysis are summarized later in section 1, and in 
section 2 a powerful RG technique is developed that is valid correct 
to fourth order in the perturbation amplitude. A technique for resolving 
the resonance structure of the Henon map is discussed in section 3, and 
in section 4 illustrative numerical results are presented. 

The present analysis assumes that a certain multipole nonlinearity is 
concentrated at a single point azimuthally located at $\theta_0$. Then 
the one-turn transfer map describing the transverse betatron motion in a 
cyclic accelerator or storage ring can be written in the form 
\cite{Tzenovb} 
\begin{equation} 
{\bf z}_{n+1} = {\widehat{\cal R}}_t 
{\left[ {\bf z}_n + {\frac {l} {R}} {\bf F} 
{\left( {\bf z}_n; \theta_0 \right)} \right]}, 
\label{Tmottransmap} 
\end{equation} 
\noindent 
where ${\bf z}$ is the state vector 
\begin{equation} 
{\bf z} = {\left( 
\begin{array}{c} 
X \\ 
P_x \\ 
Z \\ 
P_z 
\end{array} 
\right)}, 
\label{Tmstvec} 
\end{equation} 
\noindent 
${\widehat{\cal R}}_t$ is an orthogonal matrix 
\begin{equation} 
{\widehat{\cal R}}_t = {\left( 
\begin{array}{clcr} 
\cos 2 \pi \nu_x & 
\sin 2 \pi \nu_x & 0 & 0 \\ 
- \sin 2 \pi \nu_x 
& \cos 2 \pi \nu_x & 0 & 0 \\ 
0 & 0 & \cos 2 \pi \nu_z & 
\sin 2 \pi \nu_z \\ 
0 & 0 & - \sin 2 \pi \nu_z & 
\cos 2 \pi \nu_z
\end{array} 
\right)}, 
\label{Tmrotmatr} 
\end{equation} 
\noindent 
and ${\bf F} {\left( {\bf z}; \theta_0 \right)}$ is defined by 
\begin{equation} 
{\bf F} {\left( {\bf z}; 
\theta_0 \right)} = {\left( 
\begin{array}{c} 
0 \\ 
- {\frac {{\textstyle \partial V 
{\left( X, Z; \theta_0 \right)}}} 
{{\textstyle \partial X}}} \\ 
0 \\ 
- {\frac {{\textstyle \partial V 
{\left( X, Z; \theta_0 \right)}}} 
{{\textstyle \partial Z}}} 
\end{array} 
\right)}. 
\label{Tmkfmatr} 
\end{equation} 
\noindent 
Here, $(X, Z)$ is the transverse displacement, ${\left( P_x, P_z 
\right)}$ is the transverse canonical momentum, $l$ is the length of 
the multipole element, $R$ is the mean machine radius and $\nu_{x,z}$ 
are the horizontal and the vertical betatron tunes, respectively. 
Furthermore, $V {\left( X, Z; \theta \right)}$ is the nonlinear potential, 
which can be expressed in the polynomial form 
\begin{equation} 
V {\left( X, Z; \theta \right)} = 
\sum \limits_{I=2}^{\infty} 
\sum \limits_{{k, m = 0} 
\atop{k + m = I}}^{I} b_{km}^{(I)} 
{\left( \theta \right)} X^k Z^m, 
\label{Tmpotent} 
\end{equation} 
\noindent 
where $ b_{km}^{(I)} {\left( \theta \right)}$ are coefficients 
(generally functions of the azimuthal angle $\theta$) representing the 
strength of the nonlinearity. 

\renewcommand{\theequation}{\thesection.\arabic{equation}}

\setcounter{equation}{0}

\section{The Henon Map}

The simplest nontrivial example of a polynomial transfer map is the 
so-called Henon map \cite{Henon1,Henon2}. It can describe the horizontal 
betatron oscillations in an accelerator possessing a FODO-cell structure 
with a single thin sextupole. The two-dimensional Henon map can be 
obtained from equation (\ref{Tmottransmap}) in the case when the 
potential $V(X, Z; \theta)$ [see equation (\ref{Tmpotent})] contains a 
single localized cubic nonlinearity. In explicit form it can be written 
as 
\begin{equation} 
X_{n+1} = X_n \cos \omega + {\left( 
P_n - {\cal S} X_n^2 \right)} \sin \omega, 
\label{Tmhenonmapx} 
\end{equation} 
\begin{equation} 
P_{n+1} = - X_n \sin \omega + {\left( 
P_n - {\cal S} X_n^2 \right)} \cos \omega, 
\label{Tmhenonmapp} 
\end{equation} 
\noindent 
where 
\begin{equation} 
\omega = 2 \pi \nu, \qquad \qquad 
{\cal S} = {\frac {l \lambda_0 {\left( 
\theta_0 \right)} \beta^{3/2} {\left( 
\theta_0 \right)}} {2R^3}}. 
\label{Tmhmapconst} 
\end{equation} 

In section 1, we briefly discussed the essence of the RG method 
and emphasized its power to handle a number of problems arising in the 
theory of continuous dynamical systems. In this section, we will 
demonstrate that the RG method can be applied successfully to study 
discrete dynamical systems, and as a particular example, we consider the 
Henon map. The latter can be further simplified by eliminating the 
canonical momentum variable $P$. Multiplying equation (\ref{Tmhenonmapx}) 
by $\cos \omega$, multiplying equation (\ref{Tmhenonmapp}) by 
$- \sin \omega$, and summing the two equations, we obtain 
\begin{equation} 
X_{n+1} \cos \omega - P_{n+1} 
\sin \omega = X_n. 
\label{Tmhmbasicp} 
\end{equation} 
\noindent 
Substitution of the recursion relation (\ref{Tmhmbasicp}) into equation 
(\ref{Tmhenonmapx}) yields the second-order difference equation for $X$ 
\begin{equation} 
{\widehat{\cal L}} X_n = X_{n+1} - 
2 X_n \cos \omega + X_{n-1} = - 
\epsilon {\cal S} X_n^2 \sin \omega. 
\label{Tmhmbaseq} 
\end{equation} 
\noindent 
Here, $\epsilon$ is a formal small parameter introduced for convenience 
to take into account the perturbative character of the sextupole 
nonlinearity. 

Next we consider an asymptotic solution of the map (\ref{Tmhmbaseq}) 
for small $\epsilon$ by means of the RG method. The naive perturbation 
expansion 
\begin{equation} 
X_n = X_n^{(0)} + \epsilon X_n^{(1)} + 
\epsilon^2 X_n^{(2)} + \dots, 
\label{Tmhmnaivexp} 
\end{equation} 
\noindent 
when substituted into equation (\ref{Tmhmbaseq}), yields the 
perturbation equations order by order 
\begin{equation} 
{\widehat{\cal L}} X_n^{(0)} = 0, 
\label{Tmhmpereq0} 
\end{equation} 
\begin{equation} 
{\widehat{\cal L}} X_n^{(1)} = - 
{\cal S} X_n^{(0){\bf 2}} \sin \omega, 
\label{Tmhmpereq1} 
\end{equation} 
\begin{equation} 
{\widehat{\cal L}} X_n^{(2)} = - 2 
{\cal S} X_n^{(0)} X_n^{(1)} 
\sin \omega, 
\label{Tmhmpereq2} 
\end{equation} 
\begin{equation} 
{\widehat{\cal L}} X_n^{(3)} = - 
{\cal S} {\left( X_n^{(1){\bf 2}} + 
2 X_n^{(0)} X_n^{(2)} \right)} 
\sin \omega, 
\label{Tmhmpereq3} 
\end{equation} 
\begin{equation} 
{\widehat{\cal L}} X_n^{(4)} = - 2 
{\cal S} {\left( X_n^{(0)} X_n^{(3)} 
+ X_n^{(1)} X_n^{(2)} \right)} 
\sin \omega. 
\label{Tmhmpereq4} 
\end{equation} 
\noindent 
Solving equation (\ref{Tmhmpereq0}) for the zeroth-order contribution, 
we obtain the simple result 
\begin{equation} 
X_n^{(0)} = A e^{i \omega n} + c.c., 
\qquad \qquad 
P_n^{(0)} = iA e^{i \omega n} + c.c., 
\label{Tmhmsol0} 
\end{equation} 
\noindent 
where $A$ is a complex constant amplitude. By virtue of equation 
(\ref{Tmhmsol0}) the first-order perturbation equation (\ref{Tmhmpereq1}) 
becomes 
\begin{equation} 
{\widehat{\cal L}} X_n^{(1)} = - 
{\cal S} {\left( A^2 e^{2i \omega n} + 
2 {\left| A \right|}^2 + c.c. \right)} 
\sin \omega. 
\label{Tmhmperteq1} 
\end{equation} 
\noindent 
The solution to equation (\ref{Tmhmperteq1}) can be expressed as 
\begin{equation} 
X_n^{(1)} = - {\cal S} {\left| A \right|}^2 
\cot {\frac {\omega} {2}} + {\frac 
{\sin \omega} {2}} {\cal S}_1 A^2 
e^{2i \omega n} + c.c., 
\label{Tmhmsol1} 
\end{equation} 
\noindent 
where 
\begin{equation} 
{\cal S}_1 = {\frac {{\cal S}} 
{\cos \omega - \cos 2 \omega}}. 
\label{Tmhms1} 
\end{equation} 
\noindent 
To avoid resonant secular terms, we assume in addition that 
\begin{equation} 
\omega \neq 2k \pi, \qquad \qquad 
\omega \neq {\frac {2 \pi} {3}} + 2k \pi, 
\qquad \qquad 
\omega \neq {\frac {4 \pi} {3}} + 2k \pi, 
\label{Tmhmresonant1} 
\end{equation} 
\noindent 
where $k$ is an integer. The properties of the Henon map in the 
case where the betatron tune is near a resonance (in particular the 
$1/3$ resonance) will be considered in the section 3. 

The second-order perturbation equation (\ref{Tmhmpereq2}) becomes 
\begin{eqnarray} 
{\widehat{\cal L}} X_n^{(2)} = - 2 
{\cal F} \sin \omega {\left| A \right|}^2 A 
e^{i \omega n} - \sin^2 \omega {\cal S} 
{\cal S}_1 A^3 e^{3i \omega n} + c.c., 
\label{Tmhmaddit} 
\end{eqnarray} 
\noindent 
where 
\begin{equation} 
{\cal F} = {\cal S} {\left( {\frac 
{{\cal S}_1} {2}} \sin \omega - 
{\cal S} \cot {\frac {\omega} {2}} \right)}. 
\label{Tmhmffunc} 
\end{equation} 
\noindent 
The solution to equation (\ref{Tmhmaddit}) can be readily expressed in 
the form 
\begin{equation} 
X_n^{(2)} = in {\cal F} {\left| A \right|}^2 
A e^{i \omega n} + {\frac {\sin^2 \omega} {2}} 
{\cal S}_1 {\cal S}_2 A^3 e^{3i \omega n} + c.c., 
\label{Tmhmsol2} 
\end{equation} 
\noindent 
where 
\begin{equation} 
{\cal S}_2 = {\frac {{\cal S}} 
{\cos \omega - \cos 3 \omega}}, 
\qquad 
\omega \neq (2k+1) \pi, 
\qquad 
\omega \neq (2k+1) {\frac {\pi} {2}}. 
\label{Tmhms2} 
\end{equation} 
\noindent 
Continuing further with the third-order calculation, we note that 
\begin{eqnarray} 
{\widehat{\cal L}} X_n^{(3)} = - {\cal S} 
{\left( \Sigma_0 {\left| A \right|}^4 + 
\Sigma_2 {\left| A \right|}^2 A^2 
e^{2i \omega n} + \Sigma_4 A^4 e^{4i \omega n} 
+ c.c. \right)}\sin \omega. 
\label{Tmhmaddit1} 
\end{eqnarray} 
\noindent 
Here, the notation 
\begin{equation} 
\Sigma_0 = {\cal S}^2 \cot^2 {\frac {\omega} {2}} 
+ {\frac {{\cal S}_1^2} {2}} \sin^2 \omega, 
\label{Tmhmsigma0} 
\end{equation} 
\begin{equation} 
\Sigma_2 = {\cal S}_1 {\left( {\cal S}_2 
\sin \omega - {\cal S} 
\cot {\frac {\omega} {2}} \right)} 
\sin \omega + 2in {\cal F}, 
\label{Tmhmsigma2} 
\end{equation} 
\begin{equation} 
\Sigma_4 = {\frac {{\cal S}_1} {4}} {\left( 
{\cal S}_1 + 4 {\cal S}_2 \right)} 
\sin^2 \omega, 
\label{Tmhmsigma4} 
\end{equation} 
\noindent 
has been introduced. The solution to the third-order perturbation 
equation (\ref{Tmhmaddit1}) is found in a straightforward manner to be 
\begin{eqnarray} 
X_n^{(3)} = - {\frac {\cal S} {2}} \Sigma_0 
{\left| A \right|}^4 \cot {\frac {\omega} {2}} + 
{\left( {\cal B} + in {\cal S}_1 {\cal F} 
\sin \omega \right)} {\left| A \right|}^2 A^2 
e^{2i \omega n} 
\nonumber 
\end{eqnarray} 
\begin{equation} 
+ {\frac {\sin \omega} {2}} {\cal S}_3 
\Sigma_4 A^4 e^{4i \omega n} + c.c., 
\label{Tmhmsol3} 
\end{equation} 
\noindent 
where 
\begin{equation} 
{\cal B} = {\frac {{\cal S}_1^2} {2}} 
{\left( {\cal S}_2 \sin \omega - {\cal S} 
\cot {\frac {\omega} {2}} \right)} 
\sin^2 \omega - {\frac {{\cal S}_1^2 {\cal F}} 
{\cal S}} \sin \omega \sin 2 \omega, 
\label{Tmhmcoefb} 
\end{equation} 
\begin{equation} 
{\cal S}_3 = {\frac {{\cal S}} 
{\cos \omega - \cos 4 \omega}}, 
\qquad 
\omega \neq {\frac {2 \pi} {5}} + 2k \pi, 
\qquad 
\omega \neq {\frac {4 \pi} {5}} + 2k \pi. 
\label{Tmhms3} 
\end{equation} 
\noindent 
Finally, we retain terms proportional to the fundamental harmonic 
$e^{i \omega n}$ in the fourth-order perturbation equation, i.e., 
\begin{eqnarray} 
{\widehat{\cal L}} X_n^{(4)} = {\left( 
{\cal C} - 2in {\cal F}^2 \sin \omega 
\right)}{\left| A \right|}^4 A 
e^{i \omega n} + c.c., 
\label{Tmhmaddit2} 
\end{eqnarray} 
\noindent 
where 
\begin{equation} 
{\cal C} = -2 {\cal S} {\left( {\cal B} + 
{\frac {{\cal S}_1^2 {\cal S}_2} {4}} 
\sin^3 \omega - {\frac {{\cal S} \Sigma_0} {2}} 
\cot {\frac {\omega} {2}} \right)} \sin \omega. 
\label{Tmhmcoefc} 
\end{equation} 
\noindent 
We obtain the secular fourth-order contribution to the fundamental 
harmonic in the form 
\begin{equation} 
X_n^{(4)} = {\left( in {\cal D} - 
{\frac {n^2 {\cal F}^2} {2}} \right)} 
{\left| A \right|}^4 A e^{i \omega n} 
+ c.c., 
\label{Tmhmsol4} 
\end{equation} 
\noindent 
where the coefficient ${\cal D}$ is defined by 
\begin{equation} 
{\cal D} = - {\frac {{\cal C} + {\cal F}^2 
\cos \omega} {2 \sin \omega}}. 
\label{Tmhmcoefd} 
\end{equation} 

Close inspection of the naive perturbation solution starting from the 
second-order result (\ref{Tmhmsol2}) shows that secular terms 
(proportional to $n$, $n^2$, etc.) are present. To remove these terms, 
we define the renormalization transformation $A \rightarrow 
{\widetilde{A}} (n)$ by collecting all terms proportional to the 
fundamental harmonic $e^{i \omega n}$. This gives 
\begin{equation} 
{\widetilde{A}} (n) = {\left[ 1 + 
i \epsilon^2 n {\cal F} {\left| A 
\right|}^2 + \epsilon^4 {\left( in 
{\cal D} - {\frac {n^2 {\cal F}^2} {2}} 
\right)} {\left| A \right|}^4 \right]} A. 
\label{Tmhmrentran} 
\end{equation} 
\noindent 
Solving perturbatively equation (\ref{Tmhmrentran}) for $A$ in terms 
of ${\widetilde{A}} (n)$, we obtain 
\begin{equation} 
A = {\left[ 1 - i \epsilon^2 n {\cal F} 
{\left| {\widetilde{A}} (n) \right|}^2 + 
O {\left( \epsilon^3 \right)}
 \right]} {\widetilde{A}} (n). 
\label{Tmhmrenpert} 
\end{equation} 
\noindent 
A discrete version of the RG equation can be defined by considering 
the difference 
\begin{equation} 
{\widetilde{A}} (n+1) - {\widetilde{A}} (n) = 
i \epsilon^2 {\cal F} {\left| A \right|}^2 
A + \epsilon^4 {\left( i {\cal D} - {\frac 
{{\cal F}^2} {2}} - n {\cal F}^2 \right)} 
{\left| A \right|}^4 A. 
\label{Tmhmrgdiff} 
\end{equation} 
\noindent 
Substituting the expression for $A$ in terms of ${\widetilde{A}} (n)$ 
[see equation (\ref{Tmhmrenpert})] into the right-hand-side of 
equation (\ref{Tmhmrgdiff}), we can eliminate the secular terms 
up to $O{\left( \epsilon^4 \right)}$. This gives 
\begin{equation} 
{\widetilde{A}} (n+1) = {\left[ 1 + 
i \epsilon^2 {\cal F} {\left| {\widetilde{A}} 
(n) \right|}^2 + \epsilon^4 {\left( i {\cal D} 
- {\frac {{\cal F}^2} {2}} \right)} {\left| 
{\widetilde{A}} (n) \right|}^4 \right]} 
{\widetilde{A}} (n). 
\label{Tmhmrgequat} 
\end{equation} 
\noindent 
This naive RG map does not preserve the symplectic symmetry of the 
original system and does not have a {\it constant of the motion}. To 
recover the symplectic symmetry, we regularize \cite{Goto} the naive 
RG map (\ref{Tmhmrgequat}) by noting that the coefficient in the square 
brackets multiplying ${\widetilde{A}} (n)$ can be exponentiated as 
\begin{equation} 
{\widetilde{A}} (n+1) = {\widetilde{A}} (n) 
\exp {\left[ i {\widetilde{\omega}} {\left( 
{\left| {\widetilde{A}} (n) \right|} 
\right)} \right]}, 
\label{Tmhmrgmap} 
\end{equation} 
\noindent 
where 
\begin{equation} 
{\widetilde{\omega}} {\left( {\left| 
{\widetilde{A}} (n) \right|} \right)} = 
\epsilon^2 {\cal F} {\left| {\widetilde{A}} 
(n) \right|}^2 + \epsilon^4 {\cal D} {\left| 
{\widetilde{A}} (n) \right|}^4. 
\label{Tmhmtuneshift} 
\end{equation} 
\noindent 
Although the renormalization procedure described above may seem somewhat 
artificial, it holds in all orders. By extracting a symplectic implicit 
map in terms of the real part and the argument (phase) of the renormalized 
amplitude ${\widetilde{A}} (n)$, a partial proof (up to fourth order) 
of this assertion will be presented in the next paragraph. It is clear 
now that the regularized RG map (\ref{Tmhmrgmap}) possesses the obvious 
integral of motion 
\begin{equation} 
{\left| {\widetilde{A}} (n+1) \right|} = 
{\left| {\widetilde{A}} (n) \right|} = 
\sqrt{\frac {\cal J} {2}}. 
\label{Tmhmintmot} 
\end{equation} 

It is important to note that the secular terms encountered in the higher 
harmonics ${\left( e^{2i \omega n}, \; e^{3i \omega n}, \; {\rm etc}. 
\right)}$ can be summed to give the renormalized amplitudes, which 
when expressed in terms of ${\widetilde{A}} (n)$ do not contain secular 
terms. This means that once the amplitude of the fundamental harmonic 
is renormalized, any problems associated with divergences in higher 
harmonics are being remedied automatically. To demonstrate this, we 
express the amplitude of the second harmonic as 
\begin{equation} 
A_2 = \epsilon {\left[ {\frac {{\cal S}_1} {2}} 
\sin \omega + \epsilon^2 {\left( {\cal B} + 
in {\cal S}_1 {\cal F} \sin \omega \right)} 
{\left| A \right|}^2 \right]} A^2, 
\label{Tmhmsechar} 
\end{equation} 
\noindent 
which by virtue of equation (\ref{Tmhmrenpert}) acquires the form 
\begin{equation} 
A_2 = \epsilon {\left[ {\frac {{\cal S}_1} {2}} 
\sin \omega + \epsilon^2 {\cal B} {\left| 
{\widetilde{A}} (n) \right|}^2 \right]} 
{\widetilde{A}}^2 (n). 
\label{Tmhmsharen} 
\end{equation} 
\noindent 
By analogy, the amplitude of the third harmonic 
\begin{equation} 
A_3 = \epsilon^2 {\left[ {\frac {{\cal S}_1 
{\cal S}_2} {2}} \sin^2 \omega + \epsilon^2 
{\left( {\cal B}_3 + {\frac {3in} {2}} {\cal S}_1 
{\cal S}_2 {\cal F} \sin^2 \omega \right)} 
{\left| A \right|}^2 \right]} A^3 
\label{Tmhmsechar3} 
\end{equation} 
\noindent 
can also be renormalized. The result is 
\begin{equation} 
A_3 = \epsilon^2 {\left[ {\frac {{\cal S}_1 
{\cal S}_2} {2}} \sin^2 \omega + \epsilon^2 
{\cal B}_3 {\left| {\widetilde{A}} (n) \right|}^2 
\right]} {\widetilde{A}}^3 (n). 
\label{Tmhmsharen3} 
\end{equation} 

Proceeding in a manner similar to above, we can represent the canonical 
conjugate momentum $P_n$ according to 
\begin{equation} 
P_n = i {\widetilde{B}} (n) e^{i \omega n} 
+ c.c. + {\rm higher \; harmonics}, 
\label{Tmhmmoment} 
\end{equation} 
\noindent 
where 
\begin{equation} 
{\widetilde{B}} (n+1) = {\widetilde{B}} (n) 
\exp {\left[ i {\widetilde{\omega}} {\left( 
{\left| {\widetilde{A}} (n) \right|} 
\right)} \right]}. 
\label{Tmhmrgmapb} 
\end{equation} 
\noindent 
Using now the relation (\ref{Tmhmbasicp}) between the canonical conjugate 
variables ${\left( X_n, P_n \right)}$, we can express the renormalized 
amplitude ${\widetilde{B}} (n)$ in terms of ${\widetilde{A}} (n)$ as 
\begin{equation} 
{\widetilde{B}} (n) = {\frac {i {\widetilde{A}} (n)} 
{\sin \omega}} {\left[ e^{-i {\left( \omega + 
{\widetilde{\omega}} \right)}} - \cos \omega \right]}. 
\label{Tmhmampb} 
\end{equation} 
\noindent 
In addition, the sextupole nonlinearity shifts the closed orbit by the 
constant value (in normalized coordinates) 
\begin{equation} 
X_{co} = - {\frac {\epsilon {\cal S} {\cal J}} {2}} 
{\left[ 1 + {\frac {\epsilon^2} {4}} \Sigma_0 
{\cal J} + O {\left( \epsilon^3 \right)} \right]} 
\cot {\frac {\omega} {2}}, \qquad 
P_{co} = - X_{co} \tan {\frac {\omega} {2}}, 
\label{Tmhmclosedorb} 
\end{equation} 
\noindent 
which is a common property for all odd multipole nonlinearities. 

Neglecting higher harmonics and iterating the regularized RG maps 
(\ref{Tmhmrgmap}) and (\ref{Tmhmrgmapb}), the renormalized solution 
of the Henon map can be expressed as 
\begin{equation} 
X_n^{(1)} = X_{co} + \sqrt{2 {\cal J}} \cos \psi 
{\left( {\cal J}; n \right)}, 
\label{Tmhmsolx} 
\end{equation} 
\begin{equation} 
P_n^{(1)} = P_{co} + \sqrt{2 {\cal J}} {\left[ \alpha_H^{(1)} 
({\cal J}) \cos \psi {\left( {\cal J}; n \right)} - 
\beta_H^{(1)} ({\cal J}) \sin \psi {\left( {\cal J}; 
n \right)} \right]}, 
\label{Tmhmsolp} 
\end{equation} 
\noindent 
where 
\begin{equation} 
\psi {\left( {\cal J}; n \right)} = {\left[ 
\omega + {\widetilde{\omega} (\cal J)} \right]} 
n + {\widetilde{\phi}}, 
\label{Tmhmphase} 
\end{equation} 
\begin{equation} 
\alpha_H^{(1)} ({\cal J}) = {\frac{\cos \omega - 
\cos {\left[ \omega + {\widetilde{\omega} 
(\cal J)} \right]}} {\sin \omega}}, 
\quad \quad 
\beta_H^{(1)} ({\cal J}) = {\frac{\sin {\left[ 
\omega + {\widetilde{\omega} 
(\cal J)} \right]}} {\sin \omega}}. 
\label{Tmhmalphah} 
\end{equation} 
\noindent 
It is evident that the integral of motion ${\cal J}$ has the form 
of a {\it generalized Courant-Snyder invariant} \cite{Courant}, which 
can be expressed as 
\begin{equation} 
2 {\cal J} = {\left( X^{(1)} - X_{co} \right)}^2 + 
{\frac {{\left[ P^{(1)} - P_{co} - \alpha_H^{(1)} 
({\cal J}) {\left( X^{(1)} - X_{co} \right)} \right]}^2} 
{\beta_H^{(1) {\bf 2}} ({\cal J})}}. 
\label{Tmhmcsinv} 
\end{equation} 

It is important to emphasize that equation (\ref{Tmhmcsinv}) represents 
a transcendental equation for the invariant ${\cal J}$ as a function of 
the canonical variables $(X, P)$, because the coefficients $\alpha_H$ 
and $\beta_H$ depend on ${\cal J}$. Note also that the sextupole 
nonlinearity gives rise to a nonlinear tune shift ${\widetilde{\omega}}$, 
leading to the distortion of the invariant curves [circles in normalized 
phase space $(X, P)$]\footnote{Note that $\alpha_H = 0$ and $\beta_H = 1$ 
for ${\widetilde{\omega}} = 0$.} even in an approximation where only 
the contribution of the first harmonic is taken into account. Further 
distortions of the phase-space trajectories are introduced by higher 
harmonics. 

Taking into account all harmonics up to the fifth harmonic, we can 
express the renormalized fourth-order solution of the {\it Henon} map in 
the form 
\begin{equation} 
X_n = X_n^{(1)} + \sum \limits_{M=2}^{5} {\cal X}_M 
\cos M \psi {\left( {\cal J}; n \right)}, 
\label{Tmhmrfusolx} 
\end{equation} 
\begin{equation} 
P_n = P_n^{(1)} + \sum \limits_{M=2}^{5} {\cal X}_M 
{\left[ \alpha_H^{(M)} \cos M \psi {\left( {\cal J}; n 
\right)} - \beta_H^{(M)} \sin M \psi {\left( {\cal J}; 
n \right)} \right]}. 
\label{Tmhmrfusolp} 
\end{equation} 
\noindent 
The amplitudes ${\cal X}_M$ of the various harmonics are given by the 
expressions 
\begin{equation} 
{\cal X}_2 = {\frac {\epsilon {\cal J}} {2}} 
{\left( {\cal S}_1 \sin \omega + \epsilon^2 
{\cal B} {\cal J} \right)}, \qquad 
{\cal X}_4 = {\frac {1} {4}} \epsilon^3 
{\cal S}_3 \Sigma_4 {\cal J}^2 \sin \omega, 
\label{Tmhmhar24} 
\end{equation} 
\begin{equation} 
{\cal X}_3 = {\frac {\epsilon^2 {\cal J} 
{\sqrt{\cal J}}} {2 {\sqrt{2}}}} {\left( 
{\cal S}_1 {\cal S}_2 \sin^2 \omega + \epsilon^2 
{\cal B}_3 {\cal J} \right)}, \qquad 
{\cal X}_5 = {\frac {1} {2 {\sqrt{2}}}} \epsilon^4 
{\cal C}_5 {\cal J}^2 {\sqrt{\cal J}}. 
\label{Tmhmhar35} 
\end{equation} 
\noindent 
Furthermore, similar to equation (\ref{Tmhmalphah}), the generalized 
$\alpha_H^{(M)}$ and $\beta_H^{(M)}$ functions can be expressed as 
\begin{equation} 
\alpha_H^{(M)} ({\cal J}) = {\frac{\cos \omega - 
\cos M {\left[ \omega + {\widetilde{\omega} 
(\cal J)} \right]}} {\sin \omega}}, 
\qquad 
\beta_H^{(M)} ({\cal J}) = {\frac{\sin M {\left[ 
\omega + {\widetilde{\omega} 
(\cal J)} \right]}} {\sin \omega}}. 
\label{TmhmalphahM} 
\end{equation} 

\renewcommand{\theequation}{\thesection.\arabic{equation}}

\setcounter{equation}{0}

\section{Resonance Structure of the Henon Map}

The solution to the first-order perturbation equation (\ref{Tmhmperteq1}) 
was obtained in the form (\ref{Tmhmsol1}) assuming that the unperturbed 
betatron tune $\nu$ is far from the third-order resonance $3 \nu = 1$. 
It is important to study the properties of the {\it Henon} map near a 
nonlinear resonance by means of the RG method. In what follows, we 
demonstrate how the RG reduction of the {\it Henon} map works near the 
one-third resonance. A similar procedure can be performed near all other 
resonances. 

Let us expand $\omega$ according to 
\begin{equation} 
\omega = \omega_0 + \epsilon \delta_1 
+ \epsilon^2 \delta_2 + \dots, 
\qquad \qquad 
\omega_0 = {\frac {2 \pi} {3}}. 
\label{Tmhmtune} 
\end{equation} 
\noindent 
Equation (\ref{Tmhmbaseq}) can then be expressed in alternate form 
\begin{equation} 
{\widehat{\cal L}}_0 X_n = X_{n+1} - 
2 X_n \cos \omega_0 + X_{n-1} = 2 X_n 
{\left( \cos \omega - \cos \omega_0 \right)} 
- \epsilon {\cal S} X_n^2 \sin \omega. 
\label{Tmhmbequat} 
\end{equation} 
\noindent 
The perturbation expansion (\ref{Tmhmnaivexp}), when substituted into 
equation (\ref{Tmhmbequat}), yields the perturbation equations 
\begin{equation} 
{\widehat{\cal L}}_0 X_n^{(0)} = 0, 
\label{Tmhmpereseq0} 
\end{equation} 
\begin{equation} 
{\widehat{\cal L}}_0 X_n^{(1)} = - 
{\frac {\sqrt{3}} {2}} {\left( 
2 \delta_1 X_n^{(0)} + 
{\cal S} X_n^{(0){\bf 2}} \right)}, 
\label{Tmhmpereseq1} 
\end{equation} 
\begin{equation} 
{\widehat{\cal L}}_0 X_n^{(2)} = - {\sqrt{3}} 
\delta_1 X_n^{(1)} + {\left( 
{\frac {\delta_1^2} {2}} - {\sqrt{3}} 
\delta_2 \right)} X_n^{(0)} + {\cal S} 
{\left( {\frac {\delta_1} {2}} X_n^{(0){\bf 2}} 
- {\sqrt{3}} X_n^{(0)} X_n^{(1)} \right)}. 
\label{Tmhmpereseq2} 
\end{equation} 
\noindent 
Noting that $2 \omega_0 = 2 \pi - \omega_0$, equations 
(\ref{Tmhmpereseq0})-(\ref{Tmhmpereseq2}) can be solved, yielding the 
result 
\begin{equation} 
X_n^{(0)} = A e^{i \omega_0 n} + c.c., 
\qquad \qquad 
P_n^{(0)} = iA e^{i \omega_0 n} + c.c., 
\label{Tmhmresol0} 
\end{equation} 
\begin{equation} 
X_n^{(1)} = - {\frac {{\cal S} {\sqrt{3}}} {3}} 
{\left| A \right|}^2 + in {\cal G} {\left( 
A, A^{\ast} \right)} e^{i \omega_0 n} + c.c., 
\label{Tmhmresol1} 
\end{equation} 
\begin{equation} 
X_n^{(2)} = {\cal A} + in {\cal B} + 
{\left( in {\cal C} + n^2 {\cal D} \right)} 
e^{i \omega_0 n} + c.c., 
\label{Tmhmresol2} 
\end{equation} 
\noindent 
where 
\begin{equation} 
{\cal G} {\left( A, A^{\ast} \right)} = 
\delta_1 A + {\frac {{\cal S}} {2}} 
A^{\ast {\bf 2}}, 
\label{Tmhmresg} 
\end{equation} 
\begin{equation} 
{\cal A} = {\frac {2 \delta_1} {3}} 
{\cal S} {\left| A \right|}^2, 
\qquad \qquad 
{\cal B} = {\frac {{\sqrt{3}} {\cal S}} {3}} 
{\left( {\cal G}^{\ast} A - {\cal G} 
A^{\ast} \right)}, 
\label{Tmhmresab} 
\end{equation} 
\begin{equation} 
{\cal D} = {\frac {1} {2}} {\left( - \delta_1 
{\cal G} + {\cal S} {\cal G}^{\ast} A^{\ast} 
\right)}, 
\qquad \qquad 
{\cal C} = - {\frac {\Sigma + {\cal D}} 
{\sqrt{3}}}, 
\label{Tmhmrescd} 
\end{equation} 
\begin{equation} 
\Sigma = {\left( {\frac {\delta_1^2} {2}} - 
{\sqrt{3}} \delta_2 \right)} A + 
{\frac {\delta_1 {\cal S}} {2}} A^{\ast {\bf 2}} 
+ {\cal S}^2 {\left| A \right|}^2 A. 
\label{Tmhmresigma} 
\end{equation} 

Proceeding as before, we define the renormalized amplitude by 
\begin{equation} 
{\widetilde{A}} (n) = A + i \epsilon n 
{\cal G} + \epsilon^2 {\left( in {\cal C} 
+ n^2 {\cal D} \right)}. 
\label{Tmhmresramp} 
\end{equation} 
\noindent 
Taking into account the expression 
\begin{equation} 
A = {\widetilde{A}} - i \epsilon n 
{\cal G} {\left( {\widetilde{A}}, 
{\widetilde{A}}^{\ast} \right)} + 
O {\left( \epsilon^2 \right)}, 
\label{Tmhmresnram} 
\end{equation} 
\noindent 
which relates the amplitude $A$ to the renormalized amplitude 
${\widetilde{A}} (n)$, we obtain the renormalized resonant map
\begin{equation} 
{\widetilde{A}} (n+1) = {\widetilde{A}} (n) + 
i \epsilon {\widetilde{\cal G}} (n) + 
\epsilon^2 {\left[ i {\widetilde{\cal C}} (n) 
+ {\widetilde{\cal D}} (n) \right]}, 
\label{Tmhmresrmap} 
\end{equation} 
\noindent 
where 
\begin{equation} 
{\widetilde{\Lambda}} (n) = \Lambda {\left[ 
{\widetilde{A}} (n), {\widetilde{A}}^{\ast} 
(n) \right]}, 
\label{Tmhmresnot} 
\end{equation} 
\noindent 
and $\Lambda$ represents ${\cal C}$, ${\cal D}$ or ${\cal G}$. 

It is important to note that the resonant shift in the closed orbit 
is automatically renormalized, once the renormalization transformation 
$A \rightarrow {\widetilde{A}} (n)$ has been performed. The result is 
\begin{equation} 
X_{co} (n) = {\frac {\epsilon {\cal S}} {3}} 
{\left[ - {\sqrt{3}} + 2 \epsilon \delta_1 + 
O {\left( \epsilon^2 \right)} \right]} 
{\left| {\widetilde{A}} (n) \right|}^2. 
\label{Tmhmrescor} 
\end{equation} 
\noindent 
Note that the closed orbit can be corrected up to third order (in the 
sextupole strength ${\cal S}$) by choosing the first-order resonance 
detuning $\delta_1$ to be $\delta_1 = {\sqrt{3}}/2$. In terms of 
betatron tune, this implies 
\begin{equation} 
\Delta \nu = {\frac {\sqrt{3}} {4 \pi}}. 
\label{Tmhmrestcorb} 
\end{equation} 

Since the naive RG map (\ref{Tmhmresrmap}) does not preserve the 
symplectic structure of the original Henon map, an important step 
at this point consists of constructing a symplectic map in appropriate 
variables equivalent to (\ref{Tmhmresrmap}). Unfortunately, the 
regularization procedure described in the previous paragraph cannot be 
applied to the map (\ref{Tmhmresrmap}). The reason is that in the 
resonant case ${\left| {\widetilde{A}} (n) \right|}$ is no longer an 
integral of motion. An alternative way to overcome this difficulty is 
to represent ${\widetilde{A}} (n)$ as 
\begin{equation} 
{\widetilde{A}} (n) = {\sqrt{{\cal J}_n}} 
e^{i \varphi_n}, 
\label{Tmhmresacang} 
\end{equation} 
\noindent 
and attempt to find an (implicit) map in terms of the new variables 
${\left( \varphi_n, {\cal J}_n \right)}$ of the form 
\begin{equation} 
\varphi_{n+1} = \varphi_n + g {\left( 
{\cal J}_{n+1}, \varphi_n \right)}, 
\qquad \qquad 
{\cal J}_{n+1} = {\cal J}_n + f {\left( 
{\cal J}_{n+1}, \varphi_n \right)}. 
\label{Tmhmresaamap} 
\end{equation} 
\noindent 
Expanding the unknown functions $f$ and $g$ in a perturbation series 
\begin{eqnarray} 
f = \sum \limits_{k=1}^{\infty} \epsilon^k f_k, 
\qquad \qquad 
g = \sum \limits_{k=1}^{\infty} \epsilon^k g_k, 
\label{Tmhmresaddit} 
\end{eqnarray} 
\noindent 
and substituting equation (\ref{Tmhmresacang}) into equation 
(\ref{Tmhmresrmap}), we can determine $f$ and $g$ up to second order. 
We obtain 
\begin{equation} 
f_1 {\left( \varphi, {\cal J} \right)} = 
{\cal S} {\cal J}^{3/2} \sin 3 \varphi, 
\qquad \qquad 
g_1 {\left( \varphi, {\cal J} \right)} = 
\delta_1 + 
{\frac {\cal S} {2}} {\sqrt{\cal J}} 
\cos 3 \varphi, 
\label{Tmhmresfg1} 
\end{equation} 
\begin{equation} 
f_2 {\left( \varphi, {\cal J} \right)} = 
3 \delta_1 {\cal S} {\cal J}^{3/2} 
\cos 3 \varphi - {\frac {\delta_1 {\cal S} 
{\sqrt{3}}} {2}} {\cal J}^{3/2} 
\sin 3 \varphi + {\frac {3 {\cal S}^2} {4}}
{\cal J}^2 \cos 6 \varphi, 
\label{Tmhmresf2} 
\end{equation} 
\begin{eqnarray} 
g_2 {\left( \varphi, {\cal J} \right)} = 
\delta_2 - {\frac {5 {\cal S}^2 {\sqrt{3}}} 
{12}} {\cal J} - {\frac {3 \delta_1 {\cal S}} {4}} 
{\sqrt{\cal J}} \sin 3 \varphi 
\nonumber 
\end{eqnarray} 
\begin{equation} 
- {\frac {\delta_1 {\cal S} {\sqrt{3}}} {4}} 
{\sqrt{\cal J}} \cos 3 \varphi - 
{\frac {{\cal S}^2} {4}}{\cal J} \sin 6 \varphi, 
\label{Tmhmresg2} 
\end{equation} 
\noindent 
The map (\ref{Tmhmresaamap}) is symplectic provided the condition 
\begin{equation} 
{\frac {\partial g} {\partial \varphi_n}} + 
{\frac {\partial f} {\partial {\cal J}_{n+1}}} = 0 
\label{Tmhmresympc} 
\end{equation} 
\noindent 
holds. To verify equation (\ref{Tmhmresympc}), we evaluate the 
determinant of its Jacobian 
\begin{eqnarray} 
\det {\left( {\widehat{\cal J}_{\cal M}} \right)} = 
\det {\left( 
\begin{array}{clcr} 
1 + g_{\varphi_n} + g_{{\cal J}_{n+1}} 
{\frac {{\textstyle \partial {\cal J}_{n+1}}} 
{{\textstyle \partial \varphi_n}}} & 
g_{{\cal J}_{n+1}} 
{\frac {{\textstyle \partial {\cal J}_{n+1}}} 
{{\textstyle \partial {\cal J}_n}}} \\ 
f_{\varphi_n} + f_{{\cal J}_{n+1}} 
{\frac {{\textstyle \partial {\cal J}_{n+1}}} 
{{\textstyle \partial \varphi_n}}} & 1 + 
f_{{\cal J}_{n+1}} {\frac {{\textstyle \partial 
{\cal J}_{n+1}}} {{\textstyle \partial \varphi_n}}} 
\end{array} 
\right)}, 
\label{Tmhmdeterm} 
\end{eqnarray} 
\noindent 
where subscripts denote differentiation with respect to the variables 
indicated. Taking into account 
\begin{eqnarray} 
{\frac {\partial {\cal J}_{n+1}} 
{\partial \varphi_n}} = f_{\varphi_n} + 
f_{{\cal J}_{n+1}} {\frac 
{\partial {\cal J}_{n+1}} 
{\partial \varphi_n}} \quad \Longrightarrow \quad 
{\frac {\partial {\cal J}_{n+1}} 
{\partial \varphi_n}} = 
{\frac {f_{\varphi_n}} {1 - f_{{\cal J}_{n+1}}}}, 
\label{Tmhmrelation1} 
\end{eqnarray} 
\begin{eqnarray} 
{\frac {\partial {\cal J}_{n+1}} 
{\partial {\cal J}_n}} = 1 + 
f_{{\cal J}_{n+1}} {\frac 
{\partial {\cal J}_{n+1}} 
{\partial {\cal J}_n}} \quad \Longrightarrow \quad 
{\frac {\partial {\cal J}_{n+1}} 
{\partial {\cal J}_n}} = 
{\frac {1} {1 - f_{{\cal J}_{n+1}}}}, 
\label{Tmhmrelation2} 
\end{eqnarray} 
\noindent 
we obtain 
\begin{eqnarray} 
\det {\left( {\widehat{\cal J}_{\cal M}} \right)} = 
{\frac {1 + g_{\varphi_n}} {1 - f_{{\cal J}_{n+1}}}}. 
\label{Tmhmdetermfin} 
\end{eqnarray} 
\noindent 
The requirement that $\det {\left( {\widehat{\cal J}_{\cal M}} \right)} 
= 1$ leads to the condition (\ref{Tmhmresympc}). It is straightforward 
to verify that $f$ and $g$ as given by equations (\ref{Tmhmresfg1}) - 
(\ref{Tmhmresg2}) satisfy equation (\ref{Tmhmresympc}). 

The representation (\ref{Tmhmresacang}) of the renormalized amplitude 
${\widetilde{A}} (n)$ together with (\ref{Tmhmresaamap}) can be used 
as an alternate way to obtain the exponential form (\ref{Tmhmrgmap}) 
of the RG map (\ref{Tmhmrgequat}). The expansions 
\begin{eqnarray} 
f = \sum \limits_{k=1}^{\infty} \epsilon^{2k} f_{2k}, 
\qquad \qquad 
g = \sum \limits_{k=1}^{\infty} \epsilon^{2k} g_{2k}, 
\label{Tmhmexpandfg} 
\end{eqnarray} 
\noindent 
when substituted into equation (\ref{Tmhmrgequat}), after some 
straightforward algebra lead to the result 
\begin{equation} 
f_2 {\left( \varphi, {\cal J} \right)} \equiv 0, 
\qquad \qquad 
f_4 {\left( \varphi, {\cal J} \right)} \equiv 0, 
\label{Tmhmresf12} 
\end{equation} 
\begin{equation} 
g_2 {\left( \varphi, {\cal J} \right)} = 
{\cal F} {\cal J}, 
\qquad \qquad 
g_4 {\left( \varphi, {\cal J} \right)} = 
{\cal D} {\cal J}^2. 
\label{Tmhmresg12} 
\end{equation} 
\noindent 
Thus, we obtain the symplectic implicit map 
\begin{equation} 
\varphi_{n+1} = \varphi_n + \epsilon^2 {\cal F} 
{\cal J}_{n+1} + \epsilon^4 {\cal D} {\cal J}_{n+1}^2, 
\qquad \qquad 
{\cal J}_{n+1} = {\cal J}_n, 
\label{Tmhmresaargmap} 
\end{equation} 
\noindent 
which is the RG map (\ref{Tmhmrgmap}) written for the real part and 
the argument of the amplitude ${\widetilde{A}} (n)$.

\renewcommand{\theequation}{\thesection.\arabic{equation}}

\setcounter{equation}{0}

\section{Numerical Results} 

In this section we present illustrative numerical results revealing the 
stability properties of the Henon map. We use the analytical renormalized 
solution expressed by equations (\ref{Tmhmrfusolx}) - (\ref{TmhmalphahM}) 
to construct the phase portrait of the Henon map near the third-order 
resonance with $\nu = 0.323$ (also $\nu = 0.34525$), near the fourth-order 
resonance with $\nu = 0.24$, and near the fifth-order resonance with 
$\nu = 0.19$. All calculations are performed for a relatively large value 
of the sextupole strength corresponding to ${\cal S} = 0.1$. 

In Figure 1 the phase portrait of the Henon map near the third-order 
resonance with  $\nu = 0.323$ is depicted. It shows the stability region 
and the invariant curves for different values of the modulus squared of 
the renormalized amplitude ${\widetilde{A}}$ [see equation 
(\ref{Tmhmintmot})]. As the value of the invariant ${\cal J}$ increases, 
it reaches a value ${\cal J}_{max}$ above which the phase trajectories 
begin to intersect. This is due to the fact that the perturbation 
renormalization technique, valid for unperturbed betatron tunes 
sufficiently far from resonances, does not work well in this region, 
and the reduction procedure near resonances developed in section 3 
should be employed. However, the quantity ${\cal J}_{max}$ is closely 
related to the {\it dynamic aperture}. For the case where $\nu = 0.323$ 
it is approximately ${\cal J}_{max} = 3.85$. 
Figure 2 represents the phase portrait of the Henon map for 
$\nu = 0.34525$. The value of ${\cal J}_{max}$ is the same as in the 
case where  $\nu = 0.323$. 

Figures 3 and 4 show the phase portrait of the Henon map near the 
fourth-order and the fifth-order resonances where $\nu = 0.24$ and 
$\nu = 0.19$, respectively. The dynamic aperture in the case of the 
fourth-order resonance is approximately ${\cal J}_{max} = 19.01$, while 
in the case of the fifth-order resonance it is ${\cal J}_{max} = 28.5$. 

\renewcommand{\theequation}{\thesection.\arabic{equation}}

\setcounter{equation}{0}

\section{Conclusions} 

While the renormalization group method is well established in 
applications to continuous dynamical systems, the present paper 
demonstrates that the renormalization group method can also be applied 
successfully to study discrete dynamical systems. As a particular 
example, we considered the Henon map as applied to describe the 
transverse betatron oscillations in a cyclic accelerator or storage 
ring possessing a FODO-cell structure with a single thin sextupole. 
The basic equations and Henon transfer map used in the present analysis 
were summarized in section 1, and in section 2 a powerful renormalization 
group technique was developed that is valid correct to fourth order in 
the perturbation amplitude. A technique for resolving the resonance 
structure of the Henon map was discussed in section 3, and in section 
4 illustrative numerical results were presented. To the best of our 
knowledge, the present calculation represents the first successful 
application of a powerful renormalization group method to the study of 
discrete dynamical systems in a unified manner. In section 3 it was 
shown that the renormalization group map can be further expressed in 
terms of an implicit symplectic map in both cases, far from and close 
to resonances. Further applications to discrete dynamical systems will 
include generic polynomial transfer maps, the standard Chirikov-Taylor 
map, etc. 

\subsection*{Acknowledgments}

It is a pleasure to thank Prof. Alex Chao and Prof. Y. Oono for careful 
reading of the manuscript and for making valuable comments and 
suggestions. This research was supported by the U.S. Department of 
Energy. 

 

\begin{figure} 
\centerline{\epsfxsize=15cm \epsfbox{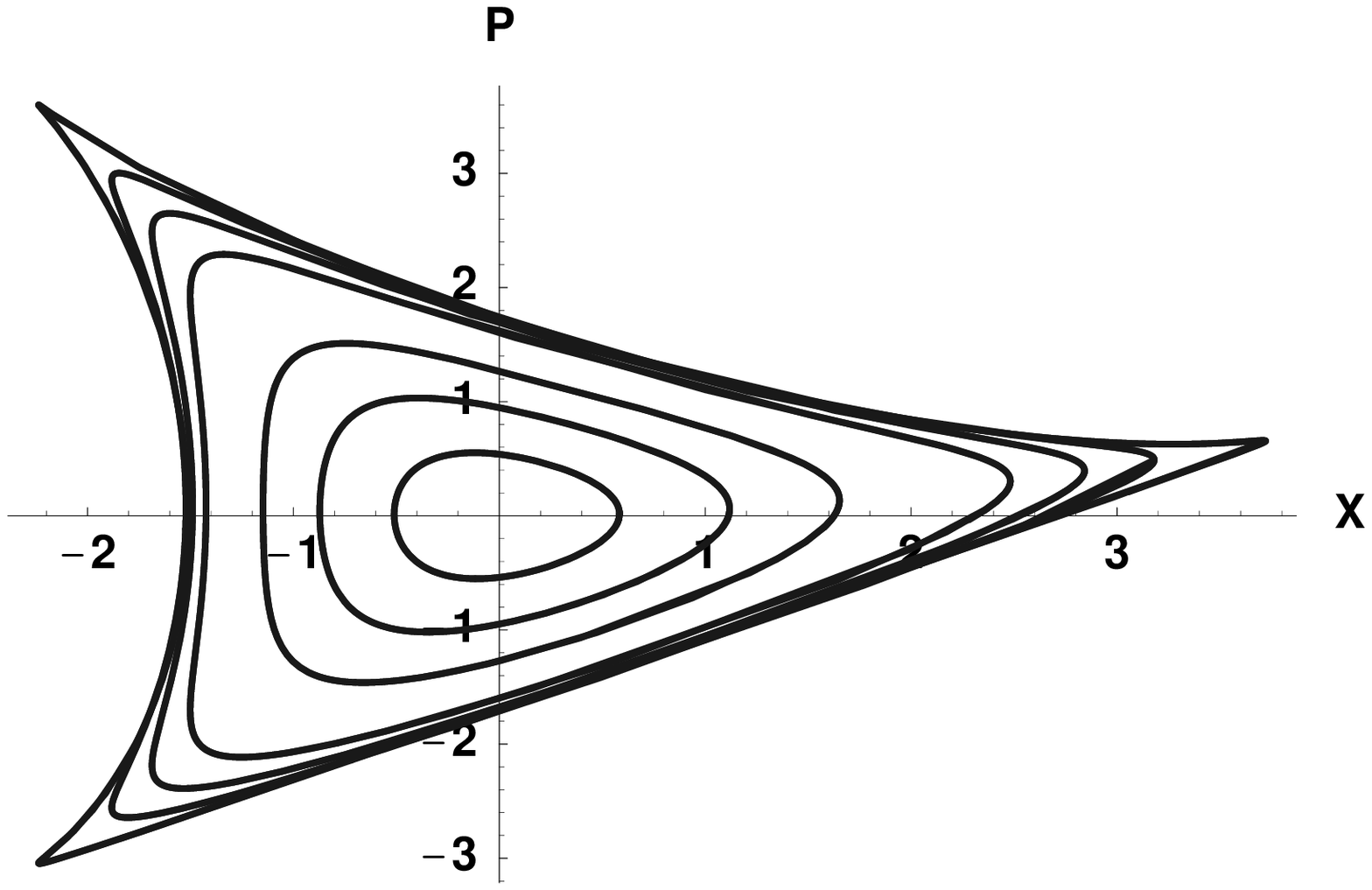}}
\caption{Phase portrait of the Henon map obtained from equations 
(\ref{Tmhmrfusolx}) - (\ref{TmhmalphahM}) near the third-order 
resonance with $\nu = 0.323$. Here, ${\cal J}$ takes values ranging 
from ${\cal J} = 0.15$ (inner contour) to ${\cal J} = {\cal J}_{max} 
= 3.85$ (outer contour).} 
\label{Fig1} 
\end{figure} 
\noindent 

\begin{figure}
\centerline{\epsfxsize=15cm \epsfbox{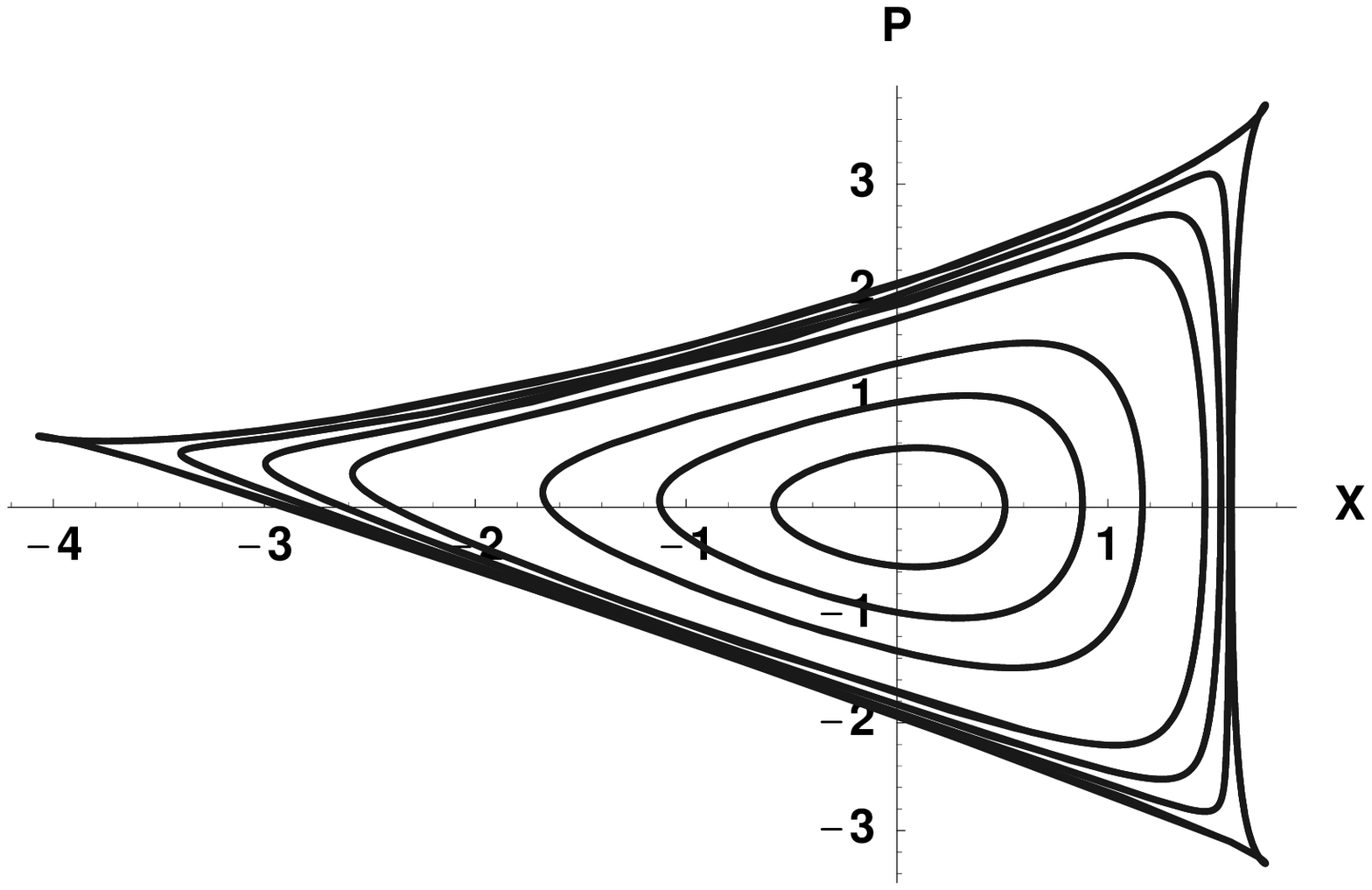}}
\caption{Phase portrait of the Henon map obtained from equations 
(\ref{Tmhmrfusolx}) - (\ref{TmhmalphahM}) near the third-order 
resonance with $\nu = 0.34525$. Here, ${\cal J}$ takes values ranging 
from ${\cal J} = 0.15$ (inner contour) to ${\cal J} = {\cal J}_{max} 
= 3.85$ (outer contour).} 
\label{Fig2} 
\end{figure}
\noindent 

\begin{figure}
\centerline{\epsfxsize=15cm \epsfbox{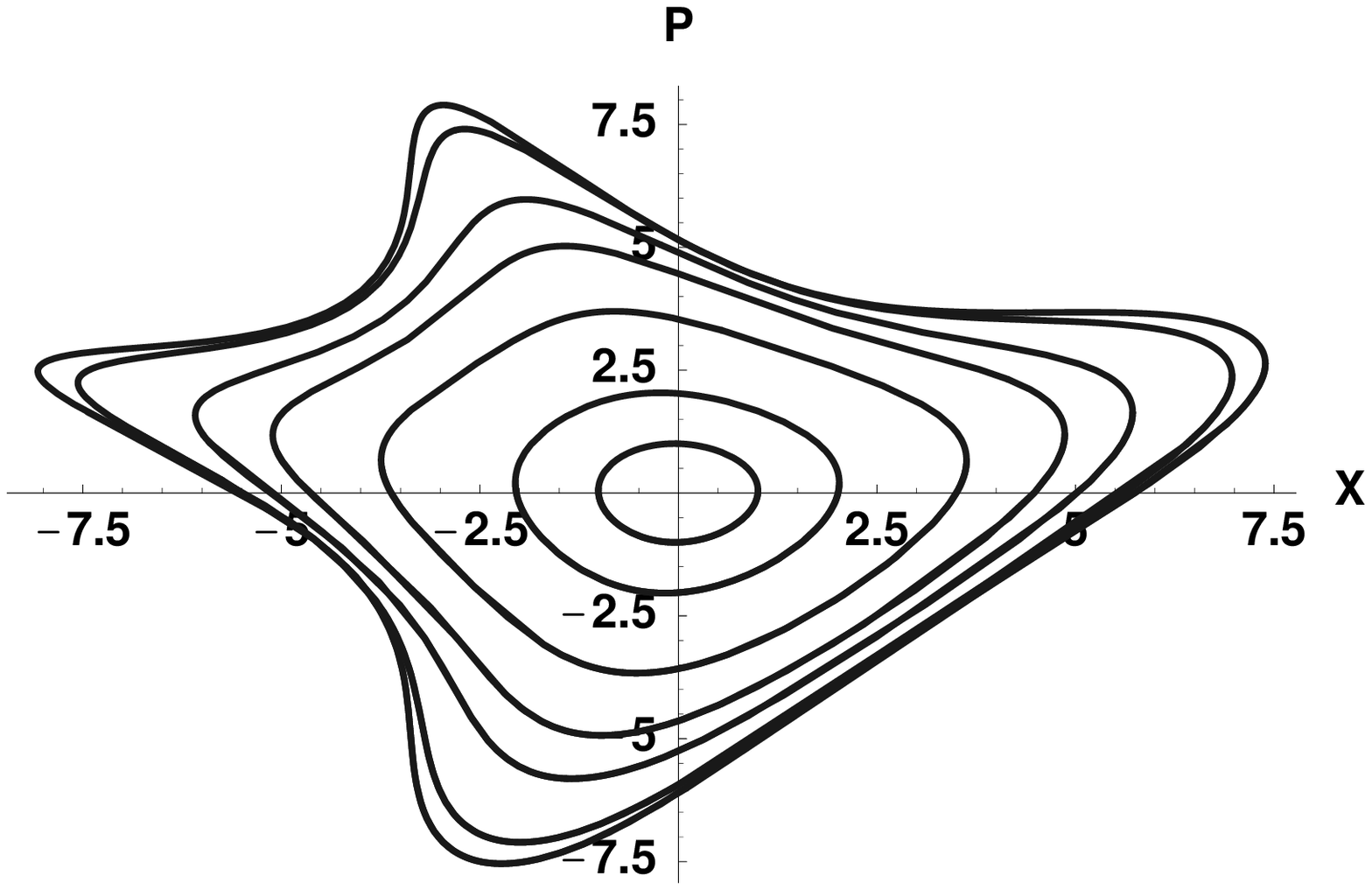}}
\caption{Phase portrait of the Henon map obtained from equations 
(\ref{Tmhmrfusolx}) - (\ref{TmhmalphahM}) near the fourth-order 
resonance with $\nu = 0.24$. Here, ${\cal J}$ takes values ranging 
from ${\cal J} = 0.5$ (inner contour) to ${\cal J} = {\cal J}_{max} 
= 19.01$ (outer contour).} 
\label{Fig3} 
\end{figure}
\noindent 

\begin{figure}
\centerline{\epsfxsize=15cm \epsfbox{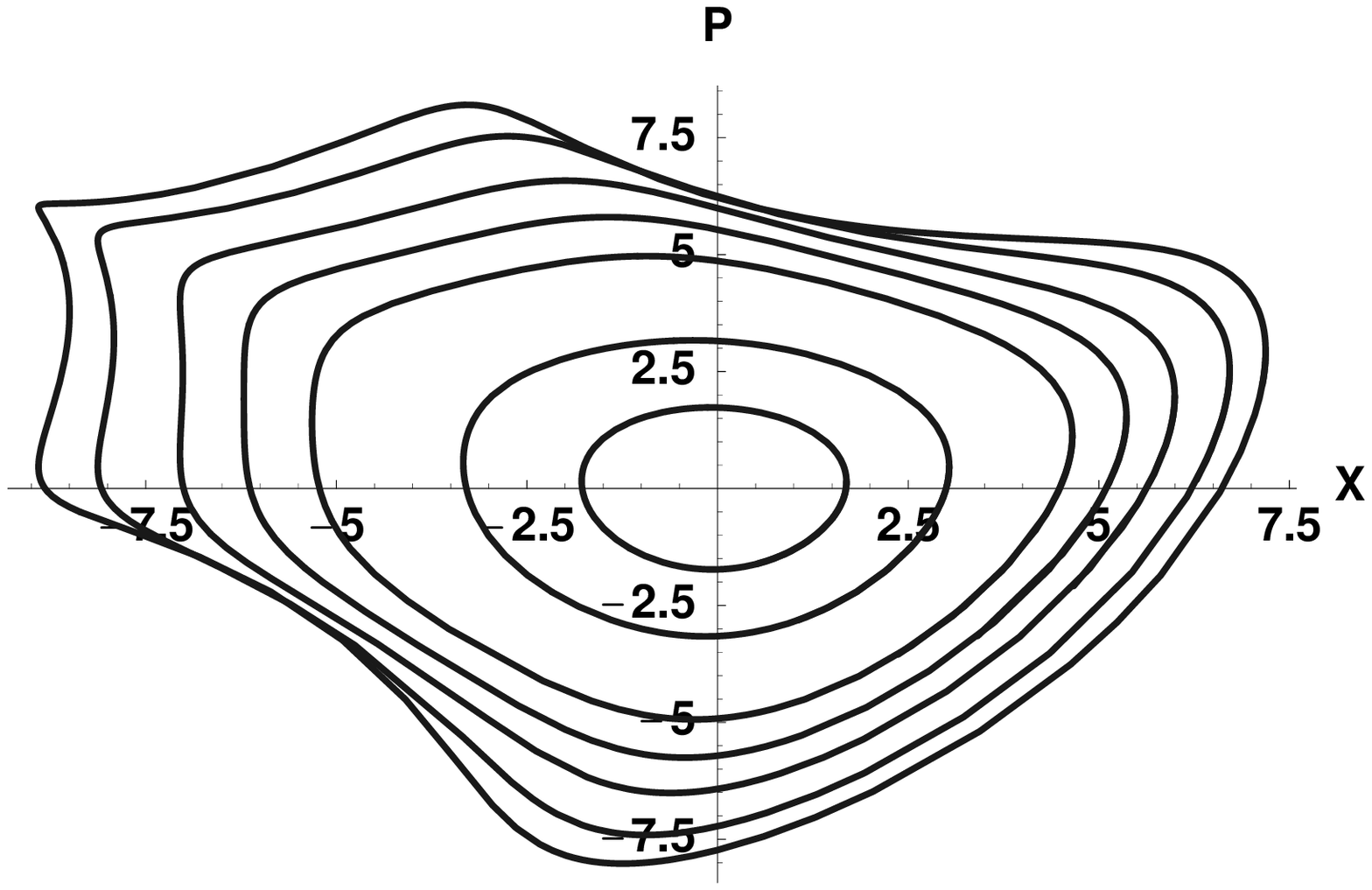}}
\caption{Phase portrait of the Henon map obtained from equations 
(\ref{Tmhmrfusolx}) - (\ref{TmhmalphahM}) near the fifth-order 
resonance with $\nu = 0.19$. Here, ${\cal J}$ takes values ranging 
from ${\cal J} = 1.5$ (inner contour) to ${\cal J} = {\cal J}_{max} 
= 28.5$ (outer contour).} 
\label{Fig4} 
\end{figure}
\noindent 


\begin{thebibliography}{9}

\bibitem{Chao} A.W. Chao, {\it ``Physics of Collective Beam Instabilities 
in High Energy Accelerators''} (John Wiley \& Sons, New York, 1993). 

\bibitem{Syphers} D.A. Edwards and M.J. Syphers, {\it ``An Introduction 
to the Physics of High Energy Accelerators''} (John Wiley \& Sons, New 
York, 1993). 

\bibitem{Davidson} R.C. Davidson and H. Qin, {\it ``Physics of Intense 
Charged Particle Beams in High Energy Accelerators''} (World Scientific, 
Singapore, 2001). 

\bibitem{Bryant} P.J. Bryant and K. Johnsen, {\it ``Principles of Circular 
Accelerators and Storage Rings''} (Cambridge University Press, 
Cambridge, 1993). 

\bibitem{Chatto1} S. Chattopadhyay {\it et al} (ed), {\it Nonlinear 
Dynamics in Particle Accelerators: Theory and Experiments (Arcidosso, 
1994) (AIP Conf. Proc. Vol. 344)} (American Institute of Physics, 
New York 1995). 

\bibitem{Chatto2} S. Chattopadhyay {\it et al} (ed), {\it Nonlinear 
and Collective Phenomena in Beam Physics (Arcidosso, 1996) (AIP Conf. 
Proc. Vol. 395)} (American Institute of Physics, 
New York 1997). 

\bibitem{Lichtenberg} A.J. Lichtenberg and M.A. Lieberman, {\it 
``Regular and Stochastic Motion''} (Springer, Berlin, 1983). 

\bibitem{Nayfeh} A.H. Nayfeh, {\it ``Introduction to Perturbation 
Techniques''} (John Wiley \& Sons, New York, 1981). 

\bibitem{Dragt} A.J. Dragt, {\it Physics of High Energy Particle 
accelerators. Lectures on Nonlinear Orbit Dynamics (Fermilab Summer 
School, 1981) (AIP Conf. Proc. Vol. 87)}, R.A. Carrigan {\it et al} 
(ed) (American Institute of Physics, New York 1982). 

\bibitem{Giacaglia} G.E.O. Giacaglia, {\it ``Perturbation Methods in 
Nonlinear Systems'' (Appl. Math. Sci. No 8)} (Springer, Berlin, 1972). 

\bibitem{Cary} J.R. Cary, {\it Phys. Rep.} {\bf 79} 129 (1981). 

\bibitem{Chen} L.-Y. Chen, N. Goldenfeld and Y. Oono, {\it Phys. 
Rev.} E{\bf 54} 376 (1996). 

\bibitem{Oono} K. Nozaki and Y. Oono, {\it Phys. Rev.} E{\bf 63} 
046101-1 (2001). 

\bibitem{Kunihiro} S.-I. Ei, K. Fujii and T. Kunihiro, {\it Ann. 
Phys., NY} {\bf 280} 236 (2000). 

\bibitem{Goto} S.-I. Goto and K. Nozaki, {\it J. Phys. Soc. Japan} 
{\bf 70} 49 (2001). 

\bibitem{Kunihir} T. Kunihiro and J. Matsukidaira, {\it Phys. Rev.} 
E{\bf 57} 4817 (1998). 

\bibitem{Arnold} V.I. Arnold and A. Avez, {\it ``Ergodic Problems of 
Classical Mechanics''} (Benjamin, New York, 1968). 

\bibitem{Zwanzig} R.W. Zwanzig, {\it Lectures in Theoretical Physics} 
Vol. 3, W.E. Brittin (ed) (John Wiley \& Sons, New York 1961). 

\bibitem{Tzenov} S.I. Tzenov, {\it New J. Phys.} {\bf 4} 6.1 (2002). 

\bibitem{Henon1} M. Henon, {\it Quart. J. Appl. Math.} {\bf 27} 291 
(1969). 

\bibitem{Henon2} M. Henon, {\it Comm. Math. Phys.} {\bf 50} 69 (1976). 

\bibitem{Tzenovb} S.I. Tzenov, {\it ``Contemporary Accelerator 
Physics''} (World Scientific, Singapore, 2003). 

\bibitem{Courant} E.D. Courant and H.S. Snyder, {\it Ann. Phys., NY} 
{\bf 3} 1 (1958). 

\end{thebibliography}
\end{document}